\documentclass[preprint,authoryear]{elsarticle}

\usepackage{amsmath}
\usepackage{graphicx}

\newtheorem{thm}{Theorem}
\newtheorem{lem}[thm]{Lemma}
\newdefinition{rmk}{Remark}
\newproof{pf}{Proof}
\newproof{pot}{Proof of Theorem \ref{thm2}}

\def\bSig\mathbf{\Sigma}

\title{Data Mining for Longitudinal Data under Multicollinearity and Time Dependence using Penalized Generalized Estimating Equations}

\author[ua]{A. Blommaert\corref{cor1}}
\ead{Adriaan.Blommaert@ua.ac.be}

\author[ua,uh]{N. Hens}
\ead{Niel.Hens@uhasselt.be}

\author[ua,sy]{Ph. Beutels}
\ead{Philippe.Beutels@ua.ac.be}

\address[ua]{Centre for Health Economics Research and Modeling Infectious Diseases,\\
	   Centre for the Evaluation of Vaccination,\\
     Vaccine and Infectious Disease Institute (WHO Collaborating Centre),\\
     University of Antwerp, Antwerp, Belgium\\
}
\address[uh]{Interuniversity Institute for Biostatistics and statistical Bioinformatics,\\
   Hasselt University \& Catholic University of Leuven, Belgium\\}
   
   \address[sy]{School of Public Health and Community Medicine,\\
  The University of New
South Wales, Sydney, Australia\\}

\cortext[cor1]{Corresponding author,\\ E-mail address: adriaan.blommaert@ua.ac.be,\\
address: Universiteitsplein 1 S4.11\\
 BE  -  2610 Wilrijk (Antwerpen), \\
 Tel:     00 323 265 29 37\\}

\begin{document}

\begin{abstract}
Penalized generalized estimating equations with Elastic Net or L2-Smoothly Clipped Absolute Deviation penalization are proposed to simultaneously 
select the most important variables and estimate their effects for 
longitudinal Gaussian data when multicollinearity is present.  The method is able to consistently select and estimate the 
main effects even when strong correlations are present. In addition, the potential pitfall of time-dependent covariates 
is clarified. Both asymptotic theory and simulation results reveal the effectiveness of penalization as a data mining 
tool for longitudinal data, especially when a large number of variables is present. The method is illustrated by mining 
for the main determinants of life expectancy in Europe.
\end{abstract}

\begin{keyword}
Covariate selection \sep Generalized estimating equations \sep Longitudinal data \sep Multicollinearity \sep Penalization \sep Time-dependent covariates
\end{keyword}

\maketitle

\section{Introduction}
\label{s:Intro}
\let\thefootnote\relax\footnote{This paper contains online supplementary material: a simulation study for binomial data}
	
Longitudinal data appear frequently in biomedical applications. Researchers are often confronted
 with the problem of determining the impact of different covariates on a response.  Correct inference 
can be obtained by building an appropriate longitudinal model. \citet{Molenberghs2005} distinguish 
three types of model families: marginal models, conditional models and subject-specific models. After the choice of the model family, an optimal set of predictors has to
 be selected. This can be a tedious task due to a large number of potential covariates.  Including irrelevant covariates leads to inefficient inference. Therefore covariate selection is an important part of longitudinal model building, which is the main focus of this paper.

Variable selection in both the mixed  model as a subject-specific model and generalized estimating equations as a marginal model will be briefly reviewed before turning to  penalization methods within the generalized estimating equations framework.

Within the mixed model framework, \citet{Wu2009} advised using significance testing or information criteria such as the Akaike information criterion or the Bayesian information criterion for the selection of fixed effects.  Information criteria have been further adapted to select both random and fixed effects in \citet{Jiang2004} and \citet{Vaida2005}. \citet{Liu1999} generalized the idea of cross-validation to mixed models. When the number of covariates becomes large, employing a stepwise search can reduce the computational burden for the selection techniques above. Recently, \citet{Jiang2008} have suggested fence methods to put up a barrier between correct and incorrect mixed models.

In this paper we choose generalized estimating equations (GEE, \citealp{Liang1986}) as our inference framework instead of generalized linear mixed models (GLMM). The GEE approach yields population averaged effects by only specifying the first two moments of the outcome distribution. Its robustness against variance structure misspecification makes the GEE method well suited for our purpose of mean structure selection. Additionally, the  problem of time-dependent covariates can be more easily clarified in the GEE context. The results provided in this paper are generalizable to linear mixed models as well, however this is not addressed here.

Despite the focus of GEE on mean structure estimation, appropriate covariate selection techniques are not well developed in this context.  The standard practice for GEE model building is stepwise selection based on Wald-type tests (see for instance \citealp{Diggle2002}). More recently some general variable selection techniques have been adapted to the GEE framework. \citet{Pan2001a} generalized the AIC to the GEE context based on the working independence assumption. \citet{Cantoni2005} suggested selection based on adequacy of prediction as measured by an adapted version of Mallow's $C_p$.  In addition to these direct methods, more computationally intensive methods  have been explored. \citet{Cantoni2007} combined cross-validation with a Markov Chain Monte Carlo based search. Alternatively, \citet{Pan2001}  proposed  minimizing a bootstrap smoothed cross-validation estimate of the expected predictive bias.

However, all of these methods lack the ability to properly deal  with a large number of covariates. Because of the discrete nature of these selection methods, the resulting estimator can become instable \citep{Breiman1996}. Moreover, complete subset comparison becomes computationally unfeasible when too many covariates are present, encouraging the use of a stepwise search. The gain in computation time by stepwise procedures comes at the  price of suboptimal prediction performance and even higher instability.

In this paper we revisit the use of penalization within the GEE context to both reduce the computational burden and tackle the problem of instability. Indeed in ordinary regression and classification problems, penalization methods are well suited and often used for the task of variable selection and regularization. The Least Adaptive Shrinkage and Selection Operator (LASSO, \citealp{Tibshirani1996}), for example, is a penalization method which achieves both subset selection and parameter shrinkage. The continuous nature of the shrinkage leads to stable selection.  The LASSO transforms the dimensionality of the subset selection problem into the selection of a single continuous tuning parameter.  A major disadvantage of the LASSO is the potentially large bias induced by its shrinkage effect. The Smoothly Clipped Absolute Deviation penalty (SCAD, Fan and Li 2001) is an adaptation to the LASSO which avoids unnecessary  bias by using a different rate of penalization depending on  the size of the coefficients. Smaller coefficients are penalized in the same manner as with the LASSO, while larger coefficients experience approximately no influence of the penalty.

Penalized generalized estimating equations (PGEE) were conceived by \citet{Fu2003} as a framework in which these penalty methods can be applied in the longitudinal context. His asymptotic results were concentrated on  bridge penalization \citep{Frank1993,Fu1998}. \citet{Dziak2006} and \citet{Dziak2007} extended this approach by using the SCAD penalty function. Even though their simulation studies display good performance of SCAD penalization for binomial data, we argue in Section~\ref{s:Asymp} that their asymptotic results are limited to the Gaussian setting.   Recently, \citet{Wang2011} have properly underpinned the SCAD penalized GEE with asymptotic theory for a response coming from the exponential family.  Moreover their asymptotic results are constructed in a high dimensional-framework, allowing for the number of covariates $p$ to diverge together with the number of clusters $n$.  Assuming only that this divergence is of the same order as the increase in the number of clusters ($p=O(n)$), whereas in other aforementioned work $p$ is assumed fixed. 

In spite of the achievements of these authors, we believe that in mining for important variables in longitudinal data, two issues are commonplace, often overlooked and could be addressed better: multicollinearity and time-dependent covariates.

In order to deal with the first issue, multicollinearity,  we suggest combining a sparse penalty function, namely the LASSO or the SCAD with a ridge part. In ordinary regression the elastic net (EN, Zou and Hastie 2005) has been proposed as the combination of the LASSO and ridge regression. Recently the SCAD penalty  has also been combined with a ridge part by \citet{Zeng2009}, an approach to which we refer hence as the SCAD$_{L2}$ penalty. The inclusion of a ridge part, adds the grouping effect to the resulting estimator. This means that highly correlated variables tend to be selected or omitted as a group.

The second issue, time-dependent covariates is often overlooked in this type of longitudinal analysis. 
Time dependence in generalized estimating equations will cause bias in the regression coefficients, unless either the cross-sectionallity assumption is satisfied or the working independence matrix is used \citep{Pepe1994,Pan2000,Diggle2002}.

In this paper we study  EN and SCAD$_{L2}$ penalization within the framework of penalized generalized estimating equations with time-dependent covariates. We show how these methods deal with selection under multicollinearity using both asymptotic theory and simulation studies. We limit ourselves to the Gaussian setting with a fixed number of covariates and present avenues for generalization to the broader exponential family. 

The remainder of the paper is organized as follows. In Section 2 we discuss the PGEE estimator with EN or SCAD$_{L2}$ penalty functions. As in \citet{Dziak2006} we establish theory by turning to the equivalent penalized generalized least squares problem, but in contrast to \citet{Dziak2006} argue that this is only possible for the Gaussian Case. The SCAD$_{L2}$ penalty function is shown to be convex under a condition on the tuning parameters. The equivalent penalized least square problem together with the convexity of the penalty function allows us to establish the grouping effect. We demonstrate how the local quadratic approximation algorithm (Fan and Li 2001) can be used to fit such a model. We also address the problem of tuning parameter selection. In Section 3 we extend the asymptotic results of \citet{Dziak2006} to EN and SCAD$_{L2}$ penalty functions. We show selection consistency, estimation consistency and asymptotic normality.  The small sample performance is investigated through simulation studies in Section 4. Section 5 presents a data example and in Section 6 we discuss our findings.

\section{Methodology}
\label{s:Meth}

\subsection{Penalized GEE}
\subsubsection{PGEE estimators}
Consider a random sample of $n$ subjects. $Y_{it}$ is the measured response for subject $i$ at time $t$ with $t=1,\ldots,T_i$. %
$\boldsymbol{X}_{it}=(X_{1,it},\ldots,X_{p,it})$ is a vector of $p$ time-dependent covariates measured at the same time as the response. Observations within a subject are correlated, observations of different subjects are assumed independent.

Without loss of generality and to facilitate the use of penalization, we assume the response to be scaled and the covariates to be standardized.

The cross-sectional influence of the covariates $\boldsymbol{X}_{it}$ on the response $Y_{it}$ is our main interest. We assume $Y$ is generated from a distribution in the exponential family with $\mbox{E}(Y_{it} |\boldsymbol{X}_{it} )=g^{-1} (\boldsymbol{X}^{T}_{it})$ , where $g$ is a known link-function.

The regression coefficients $\boldsymbol{\beta}$ can be estimated by solving the PGEE ~\citep{Fu2003}:
\begin{equation}
\label{eq:PGEE}
\boldsymbol{S^{P}}(\boldsymbol{\beta}) = \sum^{n}_{i=1}{\boldsymbol{D}^{T}_{i} V^{-1}_{i} (\boldsymbol{Y}_i - \boldsymbol{\mu}_i)} -N\boldsymbol{\dot{P}}(\boldsymbol{\beta})=\boldsymbol{0}.
\end{equation}

With $\boldsymbol{\mu}_i=g^{-1} (\boldsymbol{X}_{it}^{T})$;
 $\boldsymbol{D}_i=\boldsymbol{D}_i(\boldsymbol{\beta})=\partial\boldsymbol{\mu}_i(\boldsymbol{\beta})/\partial\boldsymbol{\beta}$;  $V_i = U_i^{1/2} W(\boldsymbol{\alpha}) U_i^{1/2}$ the working covariance matrix; $W(\boldsymbol{\alpha})$ is the working correlation matrix, parameterized with parameter vector $\boldsymbol{\alpha}$, $U_i$ is a diagonal matrix with diagonal elements $\mbox{Var}(Y_{it} |\boldsymbol{X}_{it})$; $\boldsymbol{\dot{P}}(\boldsymbol{\beta}) = \partial{P}(\boldsymbol{\beta})/\partial\boldsymbol{\beta}$ is the vector derivative of the penalty function. ~\citet{Fu2003} proposed using the bridge penalty: $ P(\boldsymbol{\beta}) = \lambda\sum\left|\beta_j\right|^\gamma $    with $\lambda\geq1$. The bridge penalty reduces to the LASSO
 when $\gamma = 1 $ and to the ridge penalty when $\gamma = 2$. Both have interesting
 properties. The first possesses the sparsity property, which implies irrelevant parameters can be set to zero. The second yields good predictive performance under multicollinearity by the grouping effect, meaning coefficients of correlated covariates tend to be equal. When $\gamma$ is chosen between one and two, the grouping effect holds, the sparsity property in contrast is lost. 
 
 We propose PGEE with a penalty function combining both the sparsity property and the grouping effect:
\begin{equation}
\label{eq:2Pen}
P(\boldsymbol{\beta})=\lambda_1P_{L1}(\boldsymbol{\beta}) + \lambda_2P_{L2}(\boldsymbol{\beta}).
\end{equation}

\subsubsection{Sparsity}

$P_{L1}$ is the part of the penalty function that provides sparsity to the resulting estimator. We explore two possibilities. The LASSO, which uses the $L1$-norm,
\[
 P_{L1}(\boldsymbol{\beta})=\sum_{j=1}^{p}\left|\beta_j\right|,
 \] 
 and the SCAD penalty: 
\[
P_{L1}(\boldsymbol{\beta})=\sum_{j=1}^{p}f_{SCAD}(\left|\beta_j\right|);
\]
\[
 f_{SCAD}(\theta)=\int^{\theta}_{0}\left\{I\left(\theta \leq \lambda_1\right) + \frac{(a\lambda_1-\theta)_{+}}{(a-1)\lambda_1} I(\theta >  \lambda_1)    \right\} d \theta,
 \] 
with $a>2$, $I(\cdot)$ the indicator function: $I(c)=1$ if $c$ is true and 0 if $c$ is false; $x_{+}= xI(x>0)$. Both provide the the sparsity property to the proposed PGEE estimator, because these penalties are singular at the origin (see discussion on sparsity in Fan and Li, 2001). The SCAD penalty behaves like the LASSO for small coefficients, whereas the estimator remains approximately unbiased for larger coefficients, because the penalty $f_{SCAD}(\theta)$ is increasing with $ \theta$ and bounded by a constant.

\subsubsection{Grouping effect}

$P_{L2}(\boldsymbol{\beta}) = \sum^{p}_{j=1}{\beta_j^2}$ is the ridge part of the penalty function. This provides the grouping effect to the resulting estimator, meaning regression coefficients of highly correlated variables tend to be equal \citep{Zou2005}. 

\begin{thm}[Grouping effect]
\label{the:Grouping}
If for all subjects $i$ the observed covariate vector $\boldsymbol{x}_{i,l}=\boldsymbol{x}_{i,k}$ , with  $l,k \in \left\{1, \ldots ,p\right\}$, then $\hat{\beta}_l= \hat{\beta}_k$, with $\boldsymbol{\hat{\beta}} = (\hat{\beta}_1, \ldots, \hat{\beta}_p)$ the solution of the PGEE (\ref{eq:PGEE}) with EN or SCAD$_{L2}$ penalty.
\end{thm}

The grouping effect property is a direct consequence of the convexity of the complete penalty function \ref{eq:2Pen}, this is satisfied if for the EN if $\lambda_2 > 0$; for the SCAD$_{L2}$ this requires $\lambda_2 > \frac{1}{2(a-1)}$. For a proof of the grouping effect we refer to Appendix A.

In summary, we propose to use the EN $($combination of LASSO and RIDGE$)$ and the SCAD$_{L2}$ $($combination of SCAD and RIDGE$)$ as the penalty in equation (\ref{eq:PGEE}). 

\subsubsection{Non-naive estimator}

~\citet{Zou2005}  show empirically that the estimator in equation (\ref{eq:PGEE}) with penalty function (\ref{eq:2Pen}) called the naive estimator  can be improved by removing the bias caused by the ridge part. 
Penalization methods aim at improving prediction performance at a cost of little extra bias
for a lower variance. By doing simulations we also observed, that a more beneficial bias-variance trade-off can be realized by removing the ridge-shrinkage by multiplying with the ridge shrinkage factor to get the non-naive PGEE estimator: 
\[
\boldsymbol{\hat{\beta}}_\text{non-naive} = \boldsymbol{\hat{\beta}}_\text{naive}*(1+\lambda_2).
\]
We will use this non-naive versus in simulation studies hence called the PGEE estimator.

\subsubsection{Time dependence}

In Section 3 we will establish asymptotic properties of the PGEE. These results are only valid if the underlying GEE estimator is consistent. The consistency property of the GEE estimator relies on the estimating equation to be unbiased: $\mbox{E}\left(\boldsymbol{S}(\boldsymbol{\beta})\right)=0$. \citet{Pepe1994} showed that when time-dependent covariates are present, this is not satisfied unless either the full covariate conditional mean (FCCM) assumption is satisfied 
\begin{equation}
\label{eq:FCCM}
\mu_{it} = \mbox{E}(Y_{it}|\boldsymbol{X}_{it}) = \mbox{E}(Y_{it}|\boldsymbol{X}_{it}\; \boldsymbol{X}_{ij}, j \neq t)
\end{equation}
or the identity working correlation matrix is used: $W(\boldsymbol{\alpha})=I$.

Using the working independence matrix, which we propose, is a correct inference tool to assess cross-sectional associations, if the FCCM-condition is not met. No additional assumptions are required besides the implied mean structure being correctly  specified: $\mu_{it}=\mbox{E}(Y_{it}|\boldsymbol{X}_{it})$. Nonetheless efficiency can be gained by using all required correct lagged covariates with another working correlation matrix. 

\subsection{Algorithm}
\label{Algorithm}

Equation (\ref{eq:PGEE}) can be solved with the local quadratic approximation (LQA) algorithm of \citet{Fan2001}. 
 We start with an initial estimate $\boldsymbol{\beta}_0=(\beta_{1,0}, \ldots, \beta_{p,0}) $ close to the solution. We thereafter iterate the following algorithm,  whereby
 $\boldsymbol{\beta}_t=(\beta_{1,t}, \ldots, \beta_{p,t})$ expresses the parameter estimate at each iteration step $t$.
 \\

STEP 1 (Remove small values):
 
If a $\beta_{j,t}$ is close to zero (closer than a predefined threshold), we set $\hat{\beta}_{j,t}=0$ and remove these variables from the model. 
\\

STEP 2 (Quadratic approximation): 
 
We approximate the derivative of the penalty
$\boldsymbol{\dot{P}}(\boldsymbol{\beta})= \left(\dot{P}(\left|\beta_1\right|),\ldots,\dot{P}(\left|\beta_p\right|)\right) $ as:
\begin{equation}
\label{eq:LQA}
\dot{P}(\left|\beta_j\right|) = \frac{\partial P(\left|\beta_j\right|)} {\partial \beta_j}
= \frac{\partial P(\left|\beta_j\right|)}{\partial \left|\beta_j\right|} 
sgn\left(\beta_j\right) \approx
\left\{   {\frac{\partial P(\left|\beta_{j,t}\right|)}{\partial\left|\beta_{j,t}\right|}} \middle/  {\left| 
\beta_{j,t} \right| } \right\} \beta_j .
\end{equation}

The penalized generalized estimating equation can hence be approximated by a Taylor series expansion of the GEE part and the approximate derivative of the penalty in (\ref{eq:LQA}):
\[
\boldsymbol{S^{P}}(\boldsymbol{\beta}) \approx \boldsymbol{S}(\boldsymbol{\beta}_t)
 + \frac{  \partial \boldsymbol{S}(\boldsymbol{\beta}_t) }   {\partial \boldsymbol{\beta}}
\left( \boldsymbol{\beta}-\boldsymbol{\beta}_t \right)
-NU\left( \boldsymbol{\beta}_t \right) - N\Sigma \left( \boldsymbol{\beta}_t \right) \left( \boldsymbol{\beta} - \boldsymbol{\beta}_t \right),
\]
with
\[
\Sigma \left( \boldsymbol{\beta_t} \right) = diag\left\{  \frac{\partial P(\left|\beta_{1,t}\right|)}{\partial \left|\beta_{1,t}\right|},\ldots,  \frac{\partial P(\left|\beta_{p,t}\right|)}{\partial \left|\beta_{p,t}\right|}
  \right\},
\]
\[
U( \boldsymbol{\beta}_t ) =  \Sigma( \boldsymbol{\beta}_t) \boldsymbol{\beta}_t.
\]
\\

STEP 3 (Beta update):
 
We update $\boldsymbol{\beta}$ as follows:
\begin{equation}
\label{eq:BetaUpdate}
\boldsymbol{\beta}_{t+1}=\boldsymbol{\beta}_t -  \left\{ \frac{  \partial \boldsymbol{S}(\boldsymbol{\beta}_t) }   {\partial \boldsymbol{\beta}} - N\Sigma \left( \boldsymbol{\beta}_t \right) \right\}^{-1}
\left\{ \boldsymbol{S}(\boldsymbol{\beta}_t) -NU\left( \boldsymbol{\beta}_t \right) \right\}, t=0, \ldots
\end{equation}

We iterate through steps 1 to 3 until convergence. Convergence is reached when the $L_2$ norm of the parameter is smaller than a predefined small constant $c$:
\[
\left\|\boldsymbol{\beta}_{t}-\boldsymbol{\beta}_{t-1}\right\|_2 < c.
\]

Notice that omitting variables in this way, has a similar drawback as backward selection: when during the iteration a parameter hits zero, it can no longer be included into the model in further iterations. The smaller the threshold, the less important this effect will be.
\citet{Hunter2005} propose a class of minorize-maximize (MM) algorithms of which the LQA algorithm is a special case. By using a perturbed version of the LQA, the drawback of not escaping zero can be avoided at the price of requiring more iterations until convergence.

A comparison of different possible algorithms falls outside the scope of this paper. 
We have implemented the LQA algorithm for solving PGEE-problems in the R-software. A version can be obtained from the authors upon request.

\subsection{Tuning parameter selection}

Using a sparse penalization technique strongly reduces the dimensionality of the covariate selection
 problem. Instead of comparing all subsets of variables, only a limited number of tuning parameters have to be selected. In the case of the EN penalty these are $\lambda_1$ and $\lambda_2$. In the case of the SCAD$_{L2}$ penalty an additional tuning parameter $a$ is involved. However, we fix $a=3.7$ as proposed by \citet{Fan2001}. To select the remaining tuning parameters, it is useful to reparametrize the penalty function (\ref{eq:2Pen}) as follows:
\begin{equation}
\label{eq:repar}
P(\boldsymbol{\beta})= \lambda \left\{ \alpha P_{L1}(\boldsymbol{\beta}) + (1 - \alpha) P_{L2}(\boldsymbol{\beta})\right\},
\end{equation}
with $\alpha\in\left[0,1\right]$ the control between sparsity and grouping effect and $\lambda$ the amount of penalization. 

Cross-validation (CV) as proposed in \cite{Cantoni2007} can directly be used to select the tuning parameters. The objective of cross-validation is to select these parameters such that a specific loss function of prediction is minimized $(\mbox{PL})$, for which the prediction loss of cross-validation $(\mbox{PL}_{CV})$ is an estimator.

  The $\mbox{PL}_{CV}$ is calculated by repeatedly splitting up the sample into a test and a training set, fitting a model on the training set and calculating the $\mbox{PL}$ on the test set. We then average the results.  Since we are working with repeated measures, the splits should be taken on the subject level rather than on the observation level.  If the number of subjects is small, leave-one-subject-out-cross-validation is a suitable choice. We use the approach proposed by \cite{Cantoni2007}:
\begin{equation}
\label{eq:CV}
\mbox{PL}_{CV} = \sum^{n}_{i=1}{\frac{(\boldsymbol{y}_i-\hat{\boldsymbol{y}}_i^{[-i]})^T V_{i}^{-1} (\boldsymbol{y}_i-\hat{\boldsymbol{y}}_i^{[-i]})  }
{T_i}},
\end{equation}
with $\hat{\boldsymbol{y}}_i^{[-i]}$ the vector of predicted responses of subject $i$ in the model fitted without this subject, $T_i$ is the number of observations in subject $i$ and $V_i$ the working covariance matrix. 
In practice the  $\mbox{PL}_{CV}$ can be minimized by a grid search. For $\lambda$ it is convenient to work at a log scale. 
 For $\alpha$ it is  sufficient to take a limited number of points within the $[0,1]$ interval. The standard error of $\mbox{PL}_{CV}$ can be calculated and used to identify a set of plausible models with a $\mbox{PL}_{CV}$ within one standard error of the optimal model.  

When one wants to avoid the the computational burden of cross-validation,  quasi-generalized cross validation (QGCV) can be 
used \citep{Fu2005}. As CV, QGCV attempts to minimize a $\mbox{PL}$ by using an estimator:  $\mbox{PL}_{QGCV}$,    
\begin{equation}
\label{eq:QGCV}
\mbox{PL}_{QGCV} = \frac{Wdev(\lambda,\alpha)} { n\left\{1-p(\lambda,\alpha)/N_{df}\right\} }.
\end{equation}

The numerator in (\ref{eq:QGCV}) is the weighted deviance: 

\[
Wdev(\lambda,\alpha)=\sum^n_{i=1}{\boldsymbol{r}_i^T R_i^{-1} \boldsymbol{r}_i}.
\] 

It  is based on the vector of deviance residuals $\boldsymbol{r}_i$ of subject $i$. The longitudinal structure is incorporated by the working correlation $R_i$. The denominator in (\ref{eq:QGCV}) is a correction for model complexity, with $N_{df}=\sum_{i=1}^n{\frac{T_i^2}{|R_i|}}$  the estimated effective number of observations and $|R_i|=\sum{\rho_{ij}}$  the sum of the elements of the working correlation matrix $R_i$. The model complexity is incorporated by an estimate of the effective number of parameters in the penalty model $p(\lambda,\alpha)$. For details see \citet{Fu2005}.

Notice that in Fu's approach a different prediction loss function is estimated, based on weighted deviance residuals instead of weighted Pearson residuals. 
In equation (\ref{eq:CV}) the ordinary difference between observed and predicted response could be replaced by the deviance residual as well, although this is not proposed in \citet{Cantoni2007}. 

Even though cross-validation is computational intensive, we recommend using this approach instead of the QGCV, because the approximation of the degrees of freedom might not be good when using the working independence matrix. Penalization paths provide an additional useful tool to
explore variable importance. The combination of cross-validation and penalization
paths is illustrated by a data example in Section \ref{s:Example}.

\section{Asymptotic theory}
\label{s:Asymp}

In this section, we establish the asymptotic properties for the naive PGEE estimator, when the number of subjects $n$ goes to infinity.  
These theorems and their proofs are adopted from ~\citet{Dziak2006}. 
Asymptotic properties are derived by turning to the equivalent penalized generalized least squares (PGLS) problem. (see Lemma~\ref{lem:PGLS}  in Appendix B). This is only possible for the Gaussian case. The reasoning behind the proofs together with a list of regularity conditions required for these proves to hold, is provided in Appendix B. Collinearity, the main context in which this type of penalization is useful, exerts no influence on these asymptotic results, but if condition
(\ref{eq:FCCM}) is not satisfied. the working independence 
correlation structure is a necessary condition for consistency to hold.

\begin{thm}[$\sqrt{n}$-consistency]
\label{th:Est_consist}
For the EN-penalty with $\lambda_1=$ $O_p(n^{-1/2})$ and $\lambda_2=$ $O_p(n^{-1/2})$, or 
for the SCAD$_{L2}$-penalty with $\lambda_1 =$ $o_p(1)$ and $\lambda_2 =$ $O_p(n^{-1/2})$, under normality and regularity conditions 1-4 in Appendix B.
there exists a sequence $\boldsymbol{\hat{\beta}}_n$ of solutions to the PGEE equation (\ref{eq:PGEE}) such that $\left\|\boldsymbol{\hat{\beta}}_n -
\boldsymbol{\beta}\right\|=$ $ O_p(n^{-1/2})$.
\end{thm}

Theorem~\ref{th:Est_consist} indicates the asymptotic estimation consistency of the PGEE estimator with EN or SCAD$_{L2}$ penalty when the number of subjects $n$ goes to infinity. This result holds if there exists a fixed true underlying vector of regression coefficients $\boldsymbol{\beta}$. The difference in conditions on the tuning parameters for the SCAD part on the one hand and the LASSO and ridge parts on the other can be attributed to the difference in the derivative of the penalty function for the non-zero regression coefficients. In order to achieve $\sqrt{n}$-consistency this rate of penalization must fade away at a high enough rate. Since the derivative of the SCAD is zero for coefficients larger than $a\lambda_1$, a moderate decrease in penalization is sufficient. 
 For asymptotic results of the SCAD penalty under a diverging number of parameters we refer to \citet{Wang2011}. Asymptotic theory in the broader context of penalized estimating function has also been described by \citet{Johnson2008}, who included results on the EN as an approximate zero-crossing. 

The aim of our proposed PGEE estimator is not only to consistently estimate the parameters of a regression model, but also to select the variables for that model. In order to get more insight in the selection properties of this estimator, we partition our vector of regression coefficients into two subgroups:
\[
\boldsymbol{\beta} =  (\boldsymbol{\beta}_\mathcal{A} , \boldsymbol{\beta}_\mathcal{N}),
\]
with $\mathcal{A}=\left\{j:\beta_j\neq0 \right\}$ the vector of indicators of non-zero coefficients, which we want to retain in our model;
with $\mathcal{N}=\left\{j:\beta_j = 0 \right\}$ the vector of indicators of zero coefficients, which we want to omit from our model. We will establish selection consistency in two steps. First we show that all active coefficients  $\boldsymbol{\beta}_\mathcal{A}$ are retained in Lemma~\ref{lem:Sens}. Second we establish conditions under which the zero coefficients  $\boldsymbol{\beta}_\mathcal{N}$ will be dropped in Lemma~\ref{lem:Sparse}. 

\begin{lem}[Sensitivity]
\label{lem:Sens}
Under the conditions in Theorem~\ref{th:Est_consist} there exists a sequence $\boldsymbol{\hat{\beta}}_n$ such that the active coefficients are included in the model with probability approaching one, i.e., $\mbox{Pr}(\exists j \in \mathcal{A}: \hat{\beta}_j=0) =o(1)$
\end{lem}

This follows directly from Theorem~\ref{th:Est_consist}. We assume that $\beta_j$ is fixed. And take some small $\epsilon > 0$. Then:
\[
\mbox{Pr}(\exists j \in \mathcal{A}: \hat{\beta}_j=0) \leq \mbox{Pr}(\exists j \in \mathcal{A}: \left|\hat{\beta}_j - \beta_j\right| > \epsilon)
\leq \mbox{Pr}(\left\|\boldsymbol{\hat{\beta}} - \boldsymbol{\beta}\right\|^2 > \epsilon^2) = o(1).
\]

For the asymptotic estimator to be sparse, we need an additional condition on the strength of tuning parameter $\lambda_1$:  
\begin{equation}
\label{eq:Sparse}
Nn^{-1/2}\lambda_1 \rightarrow \infty,
\end{equation}
as $n\rightarrow \infty$.

Let us assume the number of observations $N$ grows proportionally with the number of subjects $n$. It follows that (\ref{eq:Sparse}) is in conflict with the conditions of Theorem~\ref{th:Est_consist} for the EN, but not for the SCAD$_{L2}$.

\begin{lem}[Sparsity]
\label{lem:Sparse}
For the SCAD$_{L2}$ penalty, under conditions in Theorem~\ref{th:Est_consist} and condition (\ref{eq:Sparse}) there exists a sequence $\boldsymbol{\hat{\beta}}_n$ of solutions such that 
$\boldsymbol{\hat{\beta}}_\mathcal{A}$ is $\sqrt{n}$-consistent for $\boldsymbol{\beta}_{\mathcal{A}}$, and that 
$\boldsymbol{\hat{\beta}}_{\mathcal{N}} = \boldsymbol{\beta}_{\mathcal{N}}=\boldsymbol{0}$.
\end{lem}

\begin{thm} [Selection consistency]
\label{th:Sel_consist}
Under the conditions of Theorem~\ref{th:Est_consist} and condition (\ref{eq:Sparse}) the SCAD$_{L2}$ is asymptotically able to include the active set $(\mbox{Pr}(\exists j \in \mathcal{A}: \hat{\beta}_j=0) =o(1))$ and to omit the non active set $(\boldsymbol{\hat{\beta}}_{\mathcal{N}} = \boldsymbol{\beta}_{\mathcal{N}}=\boldsymbol{0})$. 
\end{thm}

This follows from Lemma~\ref{lem:Sens} and Lemma~\ref{lem:Sparse}.

\begin{thm}[Asymptotic normality]
Under the conditions of Theorem~\ref{th:Est_consist} there exist a sequence $\boldsymbol{\hat{\beta}}$ of solutions to equation (\ref{eq:PGEE}) such that:
\begin{equation}
\sqrt{n}\left( \boldsymbol{\hat{\beta}}_\mathcal{A} -\boldsymbol{\beta}_{\mathcal{A}}  \right)  \stackrel{L}{\rightarrow} \mbox{N}(\boldsymbol{0},\boldsymbol{\Phi}).
\end{equation}
\end{thm}
For details on $\boldsymbol{\Phi}$, see \citet{Dziak2006}.

We notice this asymptotic normality proof allows us to derive standard errors, and therefore to calculate p-values of Wald tests for parameter coefficients. However, no standard errors are provided for the parameters put to zero. Therefore we recommend using the bootstrap to obtain standard errors for all estimates.

\section{Simulation studies}
\subsection{Aims and setting}

We compare the effect of penalization on selection and prediction performance on simulated longitudinal data with multicollinearity and time-dependent covariates, using different penalty functions  (LASSO, RIDGE, EN, SCAD and SCAD$_{L2}$).  Two main settings will be investigated: 

(1) In Section~\ref{s:cross} we simulate from a cross-sectional longitudinal process and fit a PGEE model with the same time-dependent covariates. In this setting the  FCCM-assumption (see equation (\ref{eq:FCCM})) is satisfied.  

(2) In Section~\ref{s:Impl} we simulate from a process which includes lagged covariates.  The PGEE estimator is used to assess the implied cross-sectional associations between response and covariates. In both cases the working independence matrix will be used. However this requirement could be relaxed in the first setting. 

In each case the selection performance will be expressed as the percentage of correct and incorrect deletions. The prediction performance will be measured by the model error (ME, \citealp{Fan2001}):
\begin{equation}
\label{ME}
 \mbox{ME}(\hat\mu)=(\boldsymbol{\hat{\beta}}-\boldsymbol{\beta})^T \mbox{E}(XX^T)(\boldsymbol{\hat{\beta}}-\boldsymbol{\beta}).
\end{equation}

The model error is the part of the prediction error which can be influenced by the quality of the estimator $\hat{\mu}$. For a perfect estimator, this quantity is zero. The tuning parameters for the different penalty models will be selected by leave one subject out cross-validation.

\subsection{Cross-sectional model}
\label{s:cross}
We simulated longitudinal datasets with $n=20$ subjects, $T_i=T=5$ observations per subject from the following longitudinal process:
\[
Y_{it}=\boldsymbol{X}_{it}^T \boldsymbol{\beta} + e_{it}.
\]
$Y_{it}$ is the response for subject $i$ at time $t$, $\boldsymbol{\beta}=(-1,-1,1,1,0.5,0,0,0)^T$; $\boldsymbol{e}_i=(e_{i1},\ldots,{e_{iT}})^T$ multivariate normal with a first order autoregressive correlation structure with $\rho=\mbox{corr}(e_{j,it},e_{j,it+1})=0.7$. $X_{it}\sim \mbox{N}_8(0,\Sigma)$; $\Sigma$ is the correlation structure of the covariates at each time point. We use the following structure: $\mbox{corr}(X_1,X_2)=0.6$; $\mbox{corr}(X_3,X_4 )=0.3$. All other correlations between $X$-variables are set to zero.

\begin{table}
\caption{\label{t:Cross}Cross-sectionality assumption satisfied. Comparison prediction error via the ME, with Standard error (S.E.) and selection performance via ratio of correct deletions (C.D.) and ratio of Incorrect deletions (I.D.) for 100 simulations. Tuning parameters are selected using cross-validation, the median tuning parameters are displayed.}
\centering
  \begin{tabular}{l l l l l}  
  \hline
Penalty  & ME (S.E.) & ($\lambda,\alpha$) & C.D. &I.D. \\ \hline      
GEE   				&0.098 (0.006)	&	-	              & 0.000	    &0.000\\
LASSO					&0.095 (0.006)	&(0.041; 1.000)		&0.347			&0.000\\
RIDGE					&0.096 (0.005)	&(0.010; 0.000)		&0.013			&0.000\\
EN						&0.095 (0.005)	&(0.041; 0.500)		&0.333			&0.000\\
SCAD					&0.084 (0.006)	&(0.079; 1.000)		&0.630			&0.000\\
SCAD$_{L2}$		&0.081 (0.006)	&(0.170; 0.643)		&0.650			&0.002\\
\hline
\end{tabular}
\end{table}

Table \ref{t:Cross} illustrates the impact of penalization on prediction and selection performance. We notice only a small improvement in the ME by ridge, EN or LASSO penalization. The selection performance of the LASSO is slightly better than the EN, but still only removes about 35\% of variables which should be removed.

The SCAD in contrast strongly reduces the ME and achieves a good selection, considering the small sample size. With the inclusion of an additional ridge component, the SCAD$_{L2}$ accomplishes even a further reduction in the ME and a better selection, although the convexity condition ((\ref{eq:CondConvex}) in Appendix A) is not satisfied. Also in other settings, we have generally observed SCAD$_{L2}$ penalization to outperform the ordinary SCAD when collinearity is present, even if condition (\ref{eq:CondConvex}) is not fulfilled.

The large difference in the ME and selection performance for the EN and the LASSO on the one hand and the SCAD$_{L2}$  and the SCAD on the other, can likely be attributed to the combination of estimation and selection consistency (see Theorem \ref{th:Est_consist} and \ref{th:Sel_consist})  of the SCAD$_{L2}$ and the SCAD, which is lacking for the EN and the LASSO. 


\subsection{Implied cross-sectional model}
\label{s:Impl}

\subsubsection{Data generation}
\label{s:DatGen}
We make two adaptations to the simulation study in \citet{Diggle2002}. 
First we increase the number of time-dependent covariates to twenty.   Second multicollinearity is introduced by imposing a correlation structure between the covariates at each time point. The data is simulated from following process:

\[
\begin{cases}
Y_{it}= \boldsymbol{X}_{it}^T\boldsymbol{\gamma}_1 + \boldsymbol{X}_{it-1}^T\boldsymbol{\gamma}_2 + b_i + e_{it},\\
X_{j,it}= \rho_j X_{j,it-1} + \epsilon_{j,it},\text{ }\forall j=1,\ldots,20 .\\
\end{cases}
\]

Both response and covariates have a time-dependent structure. Where $i$ is the indicator of the subject, $t$ is the time indicator, and $j$ indicates the covariate. Here $e_{it} \sim \mbox{N}(0,1)$, $b_i \sim \mbox{N}(0,1) $, $\epsilon_{i0} \sim \mbox{N}(0,1)$, $\epsilon_{j,it}\sim \mbox{N}(0,1-\rho_j^2)$ for $t>0$, all mutually independent. 
$\boldsymbol{X}_{it} = (X_{1,it},\ldots, X_{20,it}) \sim \mbox{N}_{20}(0,\Sigma)$.
The covariance structure $\Sigma=diag(\Sigma_1,\Sigma_1,\Sigma_1,I_{11})$  has a block diagonal structure, with
\[ 
\Sigma_1 = \begin{bmatrix} 1&0.2&0.5 \\ 0.3&1&0.4 \\ 0.5&0.4&1 \end{bmatrix}.
\]
Notice that we consider two different types of dependence between the $X$-variables. On the one hand, each covariate for each subject $X_{j,it}$ follows its own autoregressive time evolution. On the other hand, within each subject, at each time point different covariates have a cross-sectional correlation characterized by their covariance structure $\Sigma$. 

We look at two distinct scenarios: 
\begin{itemize}
	\item Scenario 1:
	
	$\begin{cases}	
	 \boldsymbol{\gamma}_1=\boldsymbol{\gamma}_2 = \big( (2,2,2),(1,1,1),(0.1,0.1,0.1),0,0,\ldots\big), \\
	 \rho_j = 0.5, \forall j=1,\ldots,20. 
	 \end{cases}$
	
	\item Scenario 2:\\
  $\begin{cases}	
	\boldsymbol{\gamma}_1=\boldsymbol{\gamma}_2 = \big((1,1,1),(1,1,1),(0,0,0),0,0,\ldots\big),  \\
	\boldsymbol{\rho} = (0.3, 0.3, 0.3, 0.6, 0.6, 0.6, 0.5, 0.5, \ldots).
	 \end{cases}$
\end{itemize}
With $\boldsymbol{\rho}=(\rho_1, \ldots, \rho_{20})$.

For each scenario, we simulate 100 datasets with 20 subjects and 100 datasets with 100 subjects.  

\subsubsection{Cross-sectional model}

Suppose the objective is to assess the cross-sectional associations between $Y_{it}$ and $\boldsymbol{X}_{it}$. We therefore estimate the parameters of the implied cross-sectional model:   
\[
\mbox{E}(Y_{it}|\boldsymbol{X}_{it})=\boldsymbol{X}_{it}^T \boldsymbol{\beta}.
\]
The properties of the multivariate normal distribution allow us to calculate the coefficients $\boldsymbol{\beta}$ of this implied model: 
\[
\boldsymbol{\beta}=\boldsymbol{\gamma}_1 + \boldsymbol{\rho}^T \boldsymbol{\gamma}_1.
\]

\subsubsection{Simulation results}

Table~\ref{t:Impl} summarizes different PGEE fits on the 100 simulated datasets for each scenario.  SCAD$_{L2}$ penalization  yields for all scenarios the best prediction performance of all PGEE estimators. The model error is strongly reduced by using a sparse penalization technique (LASSO, EN, SCAD, SCAD$_{L2}$), especially when only 20 subjects are available. Ridge penalization in contrast improves the model error only modestly. For scenario 2 with  20 subjects, we see that the EN has a smaller model error than the SCAD. The beneficial grouping effect more than compensates for the lack of unbiasedness of the EN in this example. In general we observe the SCAD$_{L2}$ to outperform the SCAD and the EN when strong multicollinearity is present.

In addition the selection performance  of the  SCAD$_{L2}$  is superior to that of the EN or the LASSO. However, a trade-off between correct deletions and incorrect deletions is observed. In Table~\ref{t:Impl} we see that ordinary GEE and ridge penalization have sometimes a small percentage of deletions. This is due to the algorithm (see Section~\ref{Algorithm}), which includes a threshold term.

When we look at the relative bias, we observe the overall sample bias of the first beta coefficient of the  SCAD and SCAD$_{L2}$, not to be larger than that of ordinary GEE.  For the LASSO and the EN we notice a substantial negative bias.

\subsubsection{Extension to binomial data}
If we apply the logit transformation to the mean structure of these two scenarios, we can use this quantity as a probability to simulate from the Bernoulli distribution. We simulate data from following longitudinal process, which is a generalized linear mixed model:
\[
\begin{cases}
X_{j,it}= \rho_j X_{j,it-1} + \epsilon_{j,it},\text{ }\forall j=1,\ldots,20,\\
\mbox{logit}(p_{it})= \boldsymbol{X}_{it}^T\boldsymbol{\gamma}_1 + \boldsymbol{X}_{it-1}^T\boldsymbol{\gamma}_2 + b_i,\\
Y_{it} \sim Bern(p_{it}),\\
\end{cases}
\]
where $i$ indicates the subject, $t$ is the time indicator and $j$ indicates the covariate.  The response $Y_{it}$ is generated from the Bernoulli distribution, where the success probability $p_{it}$ is linked by the logit transform to a linear mean structure, with a subject-specific effect $b_i \sim \mbox{N}(0,1)$. The covariate process is autoregressive, with $\epsilon_{i0} \sim \mbox{N}(0,1)$, $\boldsymbol{X}_{it} = (X_{1,it},\ldots, X_{20,it}) \sim \mbox{N}_{20}(0,\Sigma)$. The same scenarios as for the Gaussian case were used. For details we refer to the web based supplement.

 The results of the PGEE estimator based on the binomial distribution are very similar to the ones we have displayed for the Gaussian case, indicating the suitability of these penalties outside the Gaussian context. However the improvement in the model error, even under optimal tuning are modest in comparison to the Gaussian case. For scenario 1 (100 subjects) the ordinary GEE fit displayed a model error of 0.0877 whereas the best penalization method, the SCAD$_{L2}$ had a model error of 0.0833. For more details on simulation results in the non-Gaussian setting we refer to the supplementary material to this paper.

\begin{table}
\caption{\label{t:Impl}Cross-sectionality assumption not satisfied. Comparison of the different penalty functions on the basis of  prediction performance (model error), selection performance (ratio of correct deletions (C.D.) and ratio incorrect deletions(I.D)) and relative bias of the first coefficient. Tuning parameters are selected using cross-validation, the median tuning parameters are displayed.}
\centering
\begin{tabular}{p{2cm} l l l l l} 
\hline
\multicolumn{6}{c}{Scenario 1, 20 subjects} \\
\hline
Penalty     & ME (S.E.) & ($\lambda,\alpha$)   & C.D. &I.D. & Rel. Bias \\
\hline  
GEE         &	6.489 (0.263)     &-                &	0.002	&0.000	&-0.103\\
LASSO				&4.259 (0.215)			&(0.405; 1.000)	&0.440	&0.169	&-0.158\\
RIDGE				&5.593 (0.204)			&(0.092; 0.000)	&0.002	&0.001	&-0.182\\
EN					&4.144 (0.193)			&(0.405; 0.821)	&0.392	&0.144	&-0.154\\
SCAD				&4.541 (0.287)			&(0.425; 1.000)	&0.549	&0.216	&-0.089\\
SCAD$_{L2}$	&3.744  (0.198)			&(0.518; 0.786)	&0.500	&0.184	&-0.088\\
\hline
\multicolumn{6}{c}{Scenario 1, 100 subjects} \\
\hline
GEE         &0.942 (0.027) 	&-	                &0.004	&0.001	&-0.021\\
LASSO				&0.663 (0.029)	&(0.151: 1.000)			&0.428	&0.099	&-0.047\\
RIDGE				&0.934 (0.029)	&(0.013; 0.000)			&0.011	&0.001	&-0.036\\
EN					&0.681 (0.030)	&(0.193; 0.929)			&0.417	&0.097	&-0.048\\
SCAD				&0.445 (0.026)	&(0.287; 1.000)			&0.657	&0.154	&-0.015\\
SCAD$_{L2}$	&0.438 (0.025)	&(0.349; 0.857)			&0.662	&0.158	&-0.019\\
\hline
\multicolumn{6}{c}{Scenario 2, 20 subjects} \\
  \hline
GEE         &3.018 (0.115)	&-                  &0.003	&0.000 	&-0.058\\
LASSO				&1.988 (0.100)	&(0.247; 1.000)			&0.446	&0.003	&-0.274\\
RIDGE				&2.463 (0.086)	&(0.092; 0.000)			&0.006	&0.000	&-0.246\\
EN					&1.884 (0.086)	&(0.316; 0.643)			&0.383	&0.002	&-0.261\\
SCAD				&2.280 (0.147)	&(0.287; 1.000)			&0.494	&0.023	&-0.206\\
SCAD$_{L2}$	&1.626 (0.102)	&(0.425; 0.643)			&0.431	&0.010	&-0.202\\
\hline
\multicolumn{6}{c}{Scenario 2, 100 subjects} \\
  \hline
GEE         &0.456 (0.014)	&-	              &0.003	&0.000	&-0.020\\
LASSO				&0.309 (0.014)	&(0.118; 1.000)		&0.451	&0.000	&-0.080\\
RIDGE				&0.449 (0.015)	&(0.021; 0.000)		&0.006	&0.000	&-0.046\\
EN					&0.317 (0.015)	&(0.151; 0.929)		&0.441	&0.000	&-0.086\\
SCAD				&0.161 (0.010)	&(0.235; 1.000)		&0.736 	&0.000	&-0.011\\
SCAD$_{L2}$	&0.150 (0.009)	&(0.287; 0.857)		&0.722	&0.000	&-0.029\\
\hline
\end{tabular}
\end{table}

\section{Determinants of female life expectancy in Europe}
\label{s:Example}

In health economics, identifying the most influential determinants of health-outcomes is a major research topic. If observations are taken over time, and the number of potential determinants is large, our proposed PGEE estimator can facilitate the selection of the most important ones. 
\citet{Beutels2011} for instance try to search for determinants of antibiotic consumption between European countries and over time. We will use a subset of this original dataset to search for determinants of female life expectancy. All available years between 1999 and 2005  are used in our analysis, for the following 13 countries: Austria, Belgium, Denmark, Estonia, Finland, France, Germany, Ireland, Italy, the Netherlands, Spain, Sweden and the United Kingdom. The data availability is summarized in Table \ref{t:Avail} in the Appendix. 

The objective of this study is to find the most important determinants, which could explain observed differences in female life expectancy at birth between countries and over time. A set of 42 potential determinants, having a potential direct or indirect effect on life expectancy, are included. These determinants can be roughly structured into six groups:
\textit{death rates} such as the death rate due to pneumonia; \textit{socio-economics} such as GDP per capita and hours in a work week; \textit{culture} such as the four dimensions of Hofstede: Power distance index, Individualism, Masculinity and  Uncertainty avoidance;  
factors that characterize  \textit{infectious disease preventions} such vaccination coverage against pertussis;  \textit{organization of the health care system} such as hospital beds and \textit{demographics} such as population density. 
%

These different covariates are correlated, and vary over time. The ordinary GEE estimator is unable to fit a full model, because of the limited number of data points (44) compared to the number of covariates (42). We assess the cross-sectional associations between response and covariates using the PGEE estimator. Both covariates and response are standardized prior to analysis. We employ PGEE with the penalty function parametrized as in equation (\ref{eq:repar}). Leave-one-subject-out-cross-validation is then performed on a grid of $\alpha$ and $\lambda$ values. For $\alpha$ the two extremes 0 and 1 are included. On this grid of tuning parameters the minimum of the $\mbox{PL}_{CV}$ is selected for both the EN and the SCAD$_{L2}$. For the EN, we find a $\mbox{PL}_{CV}=0.0348$ with a standard error of $0.0035$. The SCAD$_{L2}$ outperforms the EN with a $\mbox{PL}_{CV}=0.025$, with a standard error of $0.0020$. For the EN, this minimum is found for the tuning parameters $(\alpha;\lambda)=(2.857\mbox{E}-4); 3.393\mbox{E}-4)$. The SCAD$_{L2}$ reaches its optimum at  $(\alpha;\lambda)=(0.786; 0.0146)$. In both cases this minimum was significantly different from the two extreme $\alpha$ values. 
In Table~\ref{t:Fit} and \ref{t:Fit2}  the standardized coefficients of both the EN and the SCAD$_{L2}$ are reported. The EN selects 27 out of the 42, whereas the SCAD$_{L2}$ reduces this number further to 20. The order of the standardized coefficients differs only slightly between EN and SCAD$_{L2}$.

\begin{figure}
 \centering
 \includegraphics[height=10cm,width=10cm]{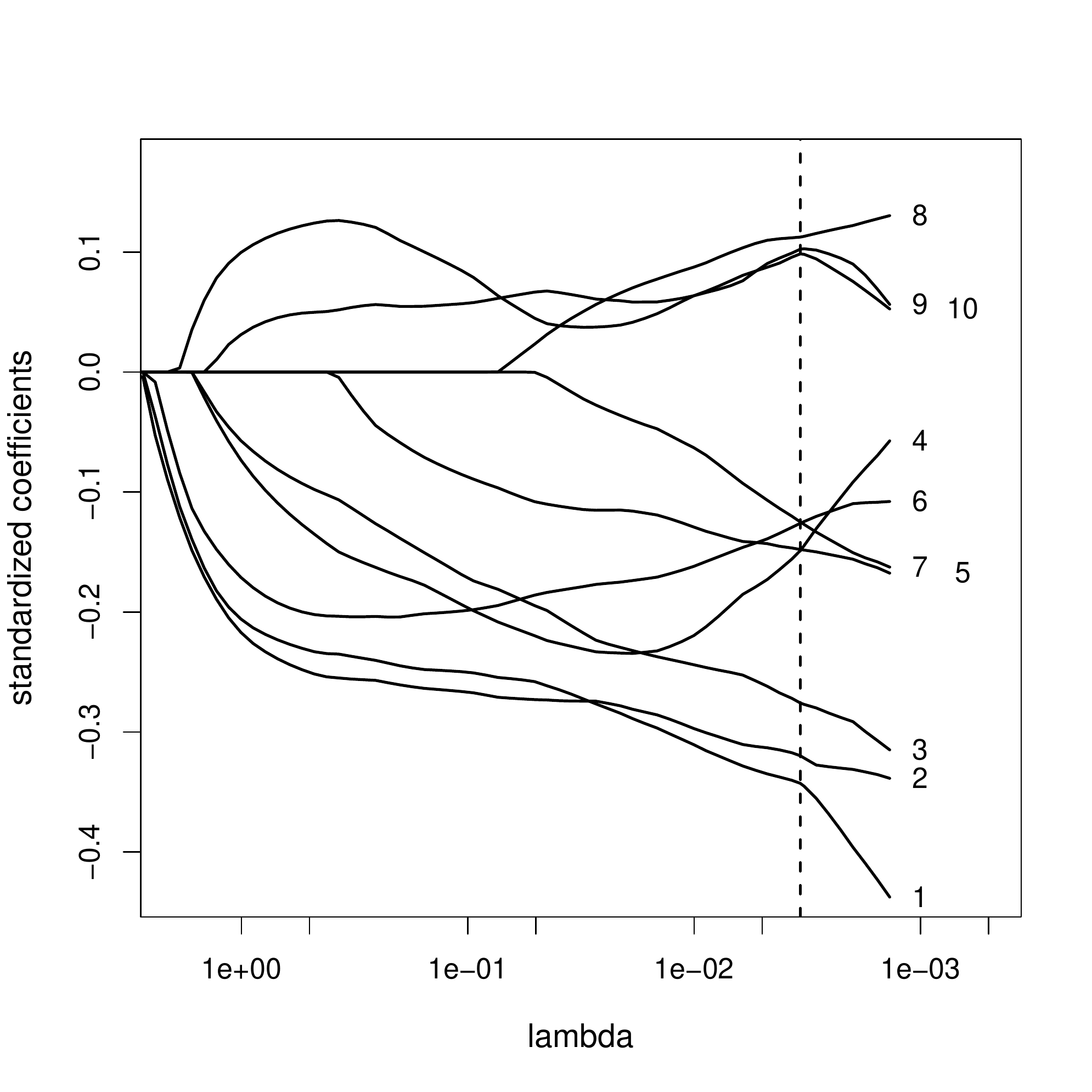}
\caption{\label{fig:path}EN plot: Standardized coefficients for the ten largest coefficients at optimal $(\alpha,\lambda)$ combination, as a function of lambda. Lambda is indicated in reverse log-scale. The coefficients are numbered by their ordering of the EN-fit as in table~\ref{t:Fit}}
\end{figure}

For EN penalization, a shrinkage plot gives some insight into the impact of the different variables on the response. Figure~\ref{fig:path} displays the shrinkage paths of the ten largest standardized coefficients at the selected fit. We keep $\alpha$  fixed and observe the evolution of standardized coefficients if $\lambda$ decreases. This shrinkage plot indicates the sensitivity of the fit with respect to $\lambda$. If $\lambda$ increases we see that some variables are excluded from the model such as the percentage children vaccinated against pertussis, whilst others increase their influence such as the infants death.

The grouping effect implies that correlated variables are pulled towards each other with increasing penalization.  The coefficients of death rates due to ischaemic heart disease and of death rate due to chronic illness, for instance, seem to converge when more penalization is performed. Our simulation studies have shown that this property results in better estimation of the regression parameters under multicollinearity, however no clear grouping structure can be identified using these methods. The shrinkage paths can at most be indicative for any grouping structure in the covariates. 

The PGEE fit provided a selection of all main determinants of female life expectancy in Europe, even in case of high correlation amongst the different covariates. In this manner, six specific death rates (due to ischaemic hearth disease; chronic diseases, pneumonia; cancer; bronchitis, asthma or emphysema; overall infant mortality) turned out to be the 'big fish' \citep{Zou2005} correlated and all predictive for lower average female life expectancy. Besides these identified death rates vaccination against pertussis, working hours per week, level of education, relative humidity and birth rate were all predictive for higher female life expectancy, while alcohol consumption and the percentage of people who believe that most people can be trusted predicted lower average female life expectancy. Penalized estimating equations can generally be applied to identify determinants for health outcomes as illustrated by this case study.

\section{Discussion}

In this paper we addressed the problem of variable selection in longitudinal data under multicollinearity and time-dependence. We proposed using the PGEE estimator under working independence to identify the most important cross-sectional associations in this context. In order to deal with potential multicollinearity, the use of a penalty function combining both the sparsity and grouping effect was proposed, namely the EN or the SCAD$_{L2}$. We proved the grouping effect as a consequence of the convexity of the penalty function and established the conditions for the SCAD$_{L2}$ to be convex. An additional consequence of this convexity is the avoidance of the multiple roots problem in penalized GEE with the SCAD penalty, mentioned in \citet{Wang2011}.

Asymptotic theory reveals that both the EN and the SCAD$_{L2}$ penalty functions can achieve a consistent fit, however only the SCAD$_{L2}$ has the additional property of selection consistency.  This is also reflected in our simulation studies were the SCAD$_{L2}$ outperformed all other methods (ordinary GEE, LASSO, ridge, EN, SCAD) with respect to prediction performance. Although the SCAD$_{L2}$ displayed overall better prediction performance this might not always be the case. In practice we recommend exploring variable importance using EN-paths and base final inference on the model selected (EN or SCAD$_{L2}$) through on cross-validation.

Combining penalty functions has proven beneficial for selection and prediction, however this requires the selection of additional tuning parameters.  In our data example we used cross-validation to select these parameters, which is computationally intensive. Establishing appropriate information criteria is crucial to the further development of penalty methods within GEE. 

Adaptive versions of the LASSO and the EN \citep{Zou2006,Zou2009} have not been considered as alternatives to the SCAD and the SCAD$_{L2}$ in this paper. In ordinary regression these penalty methods achieve selection consistency by using adaptive weights to different coefficients within the penalty function. These weights are the inverse of an initial estimate. In this manner larger initial coefficients are penalized less. 
The asymptotic theory in this paper does not incorporate adaptive versions of the penalty functions with weights dependent on the data at hand. Moreover in the presence of a large number of covariates, getting a good initial estimate is not feasible. A small simulation study (not included) suggest that the performance of the PGEE with the adaptive EN is only slightly better then the PGEE with the EN penalty, when a small number of potential covariates is present. We have also explored an adaptive SCAD$_{L2}$, but this performance was generally worse than the ordinary SCAD$_{L2}$.

In this paper, we limited ourselves to look for cross-sectional associations in the Gaussian setting with an exploration of the
binomial case. The development of asymptotic theory for the non-Gaussian setting and adaptations to the penalty function to incorporate lagged covariates are topics for future research. Asymptotic theory developed in \citet{Dziak2006} and extended in this work cannot be employed in the  non-Gaussian context because of the absence of an objective function. Asymptotic theory in the broader context of penalized estimating functions has also been described by \citet{Johnson2008}. This approach might be fruitful to derive asymptotic properties for the SCAD$_{L2}$.

\begin{table}
\caption{\label{t:Fit}Table with standardized coefficients for the PGEE estimator with both EN and SCAD$_{L2}$ penalty function. Both response and covariates were standardized before applying penalization.}
\centering
\begin{tabular}{p{5cm} l l l l l}
\hline
  &\multicolumn{2}{c}{EN fit} & \ & \multicolumn{2}{c}{SCAD$_{L2}$ fit} \\ \cline{2-3}  \cline{5-6}
  
{Variable} & {Order} & {Stand Coeff} & \ & {Order} & {Stand Coeff} \\
\hline

Death rate due to ischaemic heart disease &         				1 &     -0.343 & &         	1 &     -0.343 \\

Death rate due to chronic diseases &         								2 &     -0.320 & &         	3 &     -0.259 \\

Death rate due to pneumonia &          											3 &     -0.276 & &        	2 &     -0.284 \\

Death rate due to cancer &          												4 &     -0.149 & &         	5 &     -0.144 \\

Death rate due to bronchitis asthma \& emphysema &         	5 &     -0.148 & &         	8 &     -0.130 \\

Infant deaths per 1000 live births &          							6 &     -0.126 & &        	9 &     -0.128 \\

Percentage of urban population &          									7 &    	-0.125 & &        	4 &     -0.166 \\

Percentage of infants vaccinated against pertussis &        8 &      0.112 & &        	13 &      0.117 \\

Uncertainty Avoidance Index &          											9 &      0.102 & &         	7 &      0.136 \\

Death rate of AIDS &         																10 &      0.098 & &         6 &      0.140 \\

Relative humidity (average dew point in degrees Celsius) &        													11 &      0.094 & &        16 &      0.094 \\

Hours worked per week of full time employment &         		12 &      0.084 & &        14 &      0.097 \\

Expectancy of educational level in years &         					13 &      0.080 & &        10 &      0.123 \\

Percentage of people who believe that most people should not be trusted &   14 &      0.076 & &        17 &      0.090 \\

How strongly people find having experts, not government, make decisions for a country a bad thing  &         15 &     -0.071 & &        12 &     -0.118 \\

Birth rate &        																				16 &      0.065 & &        11 &      0.122 \\

Pure alcohol consumption, liters per capita &        				17 &     -0.057 & &        15 &     -0.096 \\
\hline
\end{tabular}
\end{table}

\begin{table}
\caption{\label{t:Fit2}Table with standardized coefficients for the PGEE estimator with both EN and SCAD$_{L2}$ penalty function (continued).  Both response and covariates were standardized before applying penalization.}
\centering
\begin{tabular}{p{5cm} l l l l l}
\hline
 &\multicolumn{2}{c}{EN fit} & \ & \multicolumn{2}{c}{SCAD$_{L2}$ fit} \\ \cline{2-3} \cline{5-6}
{Variable} & {Order}  & {Stand Coeff} & { \ } & {Order} & {Stand Coeff} \\
\hline
Women to men ratio &         																																		23 &      0.016 & &       27 &          0 \\

Private households' out-of-pocket payment on health as \% of total health expenditure &         24 &      0.012 & &        36 &          0 \\

\% of people attaining the educational level of upper secondary school &         								25 &     -0.006 & &        28 &          0 \\

Practicing physicians per 100 000 &         																										26 &     -0.004 & &        34 &          0 \\

Individualism &         																																				27 &      0.000 & &        22 &          0 \\

Masculinity &         																																					28 &          0 & &        21 &          0 \\

Power Distance Index &         																																	29 &          0 & &        23 &          0 \\

Total health expenditure, purchasing power parity in dollar per capita &         								30 &          0 & &        24 &          0 \\

Average number of people per room in an occupied housing unit &         												31 &          0 & &        25 &          0 \\

Death rate due to accidents &         																													32 &          0 & &        26 &          0 \\

Standard deviation of absolute humidity &         																							33 &          0 & &        29 &          0 \\

How strongly respect for a authority is percieved as a bad thing &        											34 &          0 & &        30 &          0 \\

How strongly atheist versus religious people describe themselves &        											35 &          0 & &        31 &          0 \\

Poverty rate &         																																					36 &          0 & &        32 &          0 \\

Average Population Density per km$^2$ &         																								37 &          0 & &        33 &          0 \\

Public sector expenditure on health as \% of total government expenditure.  &         					38 &          0 & &        35 &          0 \\

Hospital beds per 100 000 inhabitants &         																								39 &          0 & &        37 &          0 \\

Death rate due to chronic liver disease &         																							40 &          0 & &        40 &          0 \\

Death rate due to diabetes Mellitis &         																									41 &          0 & &        41 &          0 \\

Death rate due to alcohol abuse &         																											42 &          0 & &        42 &          0 \\
\hline
\end{tabular}
\end{table}

\section*{Acknowledgments}
This research was funded by the University of Antwerp's concerted research action number 23405 (BOF-GOA).
Niel Hens was funded by the UA Scientific Chair in Evidence Based Vaccinology.

\appendix

\section{Proof of the grouping effect}

We consider the PGEE as a penalized generalized least squares problem (PGLS), 
 (see Lemma~\ref{lem:PGLS}). It is then sufficient to show that the EN and SCAD$_{L2}$ penalty functions are strictly convex in $\boldsymbol{\beta}$. This is satisfied for the EN if $\lambda_2 > 0$. For the SCAD$_{L2}$ this is satisfied if $\lambda_2 > \frac{1}{2(a-1)}$ (see Lemma~\ref{lem:Convex}).

Thus choosing $\lambda_2 > \frac{1}{2(a-1)}$ yields the desired grouping effect for the SCAD$_{L2}$ PGEE estimator.
Notice that the grouping effect also holds for the non-Gaussian case with an identity working correlation matrix, because in that case, solving the PGEE is equivalent to optimizing a penalized likelihood. And the grouping effect can be proved using the same logic. 

\begin{lem}[PGEE as penalized generalized least squares]
\label{lem:PGLS}
In the Gaussian case, solving the PGEE equations is equivalent to  minimizing a penalized generalized least squares problem.
\end{lem}

\begin{pf}
Consider the following penalized generalized least squares problem: 
\begin{equation}
Q^P(\boldsymbol{\beta}) =  Q(\boldsymbol{\beta}) + NP(\boldsymbol{\beta}),
\label{eq:GLS}
\end{equation}

with:
\[
Q(\boldsymbol{\beta}) = \frac{1}{2n} \boldsymbol{S}^T K^{-1} \boldsymbol{S},
\]

\[
K= \frac{1}{n}\sum^{i=1}_{n}{\boldsymbol{D}_i^T V_i^{-1} \boldsymbol{D}_i}
= -\frac{1}{n} \frac {\partial \boldsymbol{S}(\boldsymbol{\beta})}{\partial \boldsymbol{\beta}}.
\]

This is only valid for the Gaussian case, because otherwise  $\boldsymbol{D}_i$ is a function of $\boldsymbol{\beta}$.

\[
\frac{\partial Q^p(\boldsymbol{\beta})  }
{\partial \boldsymbol{\beta}} 
= -\boldsymbol{S} + N\boldsymbol{\dot{P}}(\boldsymbol{\beta})
\]

Solving the PGEE in equation (\ref{eq:PGEE}) is hence equivalent to minimizing the objective function $Q^P(\boldsymbol{\beta})$ in equation (\ref{eq:GLS}).
\end{pf}

\begin{pf}[Theorem~\ref{the:Grouping}]
Suppose $\hat{\beta}_l  \neq \hat{\beta}_k $, let us consider $\boldsymbol{\hat{\beta^{*}}}$ as:
\[
\begin{cases}
\hat{\beta}_j = \hat{\beta}_k, & \text{if } j \neq k  \text{ and } j \neq l \\
\hat{\beta}_j^* = \frac{1}{2}(\hat{\beta}_k + \hat{\beta}_l), & \text{if } j = k  \text{ or } j = l \\
\end{cases}
\]
Suppose  $\boldsymbol{x}_{i,l} = \boldsymbol{x}_{i,k}$ for all subjects $i$, then $Q(\boldsymbol{\hat{\beta}}) = Q(\boldsymbol{\hat{\beta^*}})$,
because $X_i\boldsymbol{\hat{\beta}} = X_i\boldsymbol{\hat{\beta^*}} = \boldsymbol{D}_i, \forall \text{ subjects } i$, with $X_i=(\boldsymbol{x}_{i,1},\ldots,  \boldsymbol{x}_{i,p})$.
Since $P(\boldsymbol{\beta})$ is strictly convex, $P(\boldsymbol{\beta}) > P(\boldsymbol{\beta^*})$. Therefore
$\boldsymbol{\beta}$ cannot be the minimizer of  $Q^p(\boldsymbol{\beta})$, which is a contradiction. Therefore $\hat{\beta}_l = \hat{\beta}_k$.
\end{pf}

\begin{lem} [Convexity of the penalty function]
\label{lem:Convex}
~If $\lambda_2 > \frac{1}{2(a-1)}$, the SCAD$_{L2}$-penalty is strictly convex.
\end{lem}

\begin{pf}
Given 
\[
P(\theta)=\lambda_1 \int^{\theta}_{0}\left\{I(\theta \leq \lambda_1) + \frac{(a\lambda_1-\theta)_{+}}{(a-1)\lambda_1} I(\theta > \lambda_1)    \right\} d \theta  + \lambda_2\theta^2,
 \]

we calculate the second derivative of $P(\theta)$:

\[
\ddot{P}(\theta) =
\begin{cases}
2\lambda_2, & \text{if } \theta  \leq \lambda_1, \\
\frac{-1}{a-1} + 2\lambda_2, & \text{if } \lambda_1 < \theta \leq a\lambda_1, \\
2\lambda_2, & \text{if } a\lambda_1 < \theta.
\end{cases}
\]

This second derivative is always positive if:
\begin{equation}
\label{eq:CondConvex} 
\lambda_2 > \frac{1}{2(a-1)}.
\end{equation}
Under condition (\ref{eq:CondConvex}) the SCAD$_{L2}$ penalty function is strictly convex in the parameter vector $\boldsymbol{\beta}$.
 \end{pf}

\section{Background Asymptotic theory}
For the asymptotic properties to hold, following regularity conditions are required, which are taken over from \cite{Dziak2006}.

\vspace{3 mm}

\textbf{Regularity condition 1:}
$\boldsymbol{S}(\boldsymbol{\beta})$ and $\boldsymbol{K}(\boldsymbol{\beta}) =  \frac{1}{n}\sum^{i=1}_{n}{\boldsymbol{D}_i^T V_i^{-1} \boldsymbol{D}_i}$ have continuous third derivatives in $\boldsymbol{\beta}$.

\vspace{3 mm}

\textbf{Regularity condition 2:}
$\boldsymbol{K}(\boldsymbol{\beta})$ is positive definite with probability approaching one. there exist a non-random function $\boldsymbol{K_0}(\boldsymbol{\beta})$ such that $\left\|\boldsymbol{K}(\boldsymbol{\beta}) - \boldsymbol{K_0}(\boldsymbol{\beta})\right\| \stackrel{p}{\rightarrow} 0$ uniformly, and $\boldsymbol{K_0}(\boldsymbol{\beta}) > 0$ for all $\boldsymbol{\beta}$.

\vspace{3 mm}

\textbf{Regularity condition 3:}
The $\boldsymbol{S_i}=\boldsymbol{D}^{T}_{i} V^{-1}_{i} (\boldsymbol{Y}_i - \boldsymbol{\mu}_i)$ have finite covariance for all $  \boldsymbol{\beta} $.

\vspace{3 mm}

\textbf{Regularity condition 4:}
The derivatives of $\boldsymbol{K_0}(\boldsymbol{\beta})$ in $\boldsymbol{\beta}$ are $ O_p(1)$  for all $\boldsymbol{\beta}$.

\vspace{3 mm}

\begin{pf}[Theorem 2: $\sqrt{n}$-consistency]
~ For $0\leq\epsilon\leq1$, it is sufficient to show that with probability at least $1-\epsilon$, $\exists$ some large constant  $C_\epsilon$, such that a local solution $\hat{\boldsymbol{\beta}}_n$ of $Q^P(\boldsymbol{\beta})$ exist in the interior of the ball $
\left\{\boldsymbol{\beta} + n^{-1/2}\boldsymbol{u}: \left\|\boldsymbol{u}\right\| \leq C_\epsilon    \right\}$,

such that $\left\|\boldsymbol{\hat{\beta}}_n -
\boldsymbol{\beta}\right\| = O_p(n^{-1/2})$ will be $O_p(n^{-1/2})$, \citep{Fan2001,Dziak2006}. 
 The conditions on the tuning parameters make sure, that $Q(\boldsymbol{\beta})$ asymptotically dominates the penalty part. 
 \end{pf}
 
\begin{pf}[Lemma 2: Sparsity]
Condition (\ref{eq:Sparse}) provides that in a neighborhood where the non-active coefficients are close tho zero, the sparse part has asymptotically still enough influence to put the estimates of the non-active coefficients exactly equal to zero:
\[
Q^P(\boldsymbol{\beta}_\mathcal{A},0)=argmin\left\{Q^P(\boldsymbol{\beta}_\mathcal{A},\boldsymbol{\beta}_\mathcal{B})  \right\}
\]
\end{pf}

\section{Data availability determinants of life expectancy}

\begin{table}
\caption{\label{t:Avail}Table of availability of all covariates and response 
for the different Country-Year combinations, 1 indicates available, 0 non available}
\centering
\begin{tabular}{|l|r|r|r|r|r|r|r|}
\hline
   Country &       1999 &       2000 &       2001 &       2002 &       2003 &       2004 &       2005 \\
\hline
   Austria &          1 &          1 &          1 &          0 &          1 &          0 &          0 \\
\hline
   Belgium &          1 &          0 &          0 &          0 &          0 &          0 &          0 \\
\hline
   Denmark &          0 &          0 &          1 &          0 &          0 &          0 &          0 \\
\hline
   Estonia &          0 &          0 &          0 &          0 &          1 &          1 &          0 \\
\hline
   Finland &          1 &          1 &          1 &          1 &          0 &          1 &          1 \\
\hline
    France &          1 &          1 &          1 &          1 &          0 &          0 &          0 \\
\hline
   Germany &          1 &          1 &          1 &          0 &          0 &          0 &          0 \\
\hline
   Ireland &          1 &          1 &          1 &          0 &          1 &          1 &          0 \\
\hline
     Italy &          1 &          1 &          1 &          0 &          0 &          0 &          0 \\
\hline
Netherlands &          1 &          1 &          1 &          0 &          1 &          0 &          0 \\
\hline
     Spain &          1 &          1 &          1 &          1 &          1 &          0 &          0 \\
\hline
    Sweden &          1 &          0 &          0 &          1 &          0 &          0 &          0 \\
\hline
United Kingdom &          1 &          1 &          1 &          1 &          0 &          0 &          0 \\
\hline\end{tabular}
\end{table}

\bibliographystyle{elsarticle-harv}

\end{document}